\documentclass[11pt,a4paper]{article}
\pdfoutput=1
\usepackage{jheppub}
\usepackage{graphics}
\usepackage{stmaryrd}
\usepackage{amsfonts}
\usepackage{mathrsfs}
\usepackage{color}

\title{Nonequilibrium dynamical transition process between excited states of holographic superconductors}

\author[a,b]{Ran Li,}

\author[b,c]{Jin Wang,\footnote{Corresponding author}}

\author[d]{Yong-Qiang Wang,}

\author[e,f]{Hongbao Zhang}

\affiliation[a]{Department of Physics, Henan Normal University, Xinxiang 453007, China}

\affiliation[b]{Department of Chemistry, State University of New York at Stony Brook, Stony Brook, New York 11790, USA}

\affiliation[c]{Department of Physics and Astronomy, State University of New York at Stony Brook, Stony Brook, New York 11790, USA}

\affiliation[d]{Research Center of Gravitation and Institute of Theoretical Physics,
 Lanzhou University, Lanzhou 730000, China}

\affiliation[e]{Department of Physics, Beijing Normal University, Beijing 100875, China}

\affiliation[f]{Theoretische Natuurkunde, Vrije Universiteit Brussel, and The International Solvay
Institutes, Pleinlaan 2, B-1050 Brussels, Belgium}

\emailAdd{liran@htu.edu.cn}

\emailAdd{jin.wang.1@stonybrook.edu}

\emailAdd{yqwang@lzu.edu.cn}

\emailAdd{hzhang@vub.ac.be}

\abstract{We study the dynamics of the holographic $s$-wave superconductors described by the Einstein-Maxwell-complex scalar field theory with a negative cosmological constant. If the eigenfunction of the linearized equation of motion of the scalar field in the planar RNAdS black hole background is chosen as the initial data, the bulk system will evolve to the intermediate state that corresponds to the excited state superconductor on the boundary. The process can be regarded as the non-equilibrium condensation process of the excited state of holographic superconductor. When the linear superposition of the eigenfunctions is chosen as the initial data, the system will go through a series of the intermediate states corresponding to different overtone numbers, which can be regarded as the dynamical transition process between the excited states of holographic superconductor. Because the intermediate states are metastable, the bulk system eventually evolves to the stationary state that corresponds the ground state of the holographic superconductor. We also provide a global and physical picture of the evolution dynamics of the black hole and the corresponding superconducting phase transition from the funneled landscape view, quantifying the weights of the states and characterizing the transitions and cascades towards the ground state.}


\begin{document}

\maketitle

\section{Introduction}

The AdS/CFT correspondence \cite{Ma,GKP,Wi}, which states that the weakly coupled gravity in AdS space is dual to the strongly coupled conformal field theory on the boundary, has been applied to describe and understand the high temperature superconductivity \cite{3HPRL}. The construction of the gravitational dual of the superconductor depends mainly on the observation in \cite{Gubser} that the charged scalar perturbation in the background of the planar Reissner-Nordstrom anti-de Sitter (RNAdS) black hole below the certain critical temperature is unstable due to the violation of Breitenlohner-Freedman (BF) bound \cite{BF} in the near horizon region. When the backreaction of the scalar field  is taken into account, the hairy black hole in asymptotic AdS space with a nontrivial scalar field configuration outside of the horizon was numerically constructed as a gravity dual description of superconductivity \cite{3HJHEP}. Since then, various  holographic models have been constructed and widely studied to capture the various properties of the superconductivity with high critical temperature \cite{HLS}.

However, it should be noted that most of the work in this aspect focuses on the ground states of holographic superconductors, where the scalar fields do not have zeros (nodes) along the radial direction. Recently, inspired by the work of Gubser \cite{Gubser:2008zu}, an interesting progress has been made in Ref. \cite{ESC} where the gravitational dual of the excited states of the holographic superconductors was constructed. The excited states are characterized by the number of nodes of the scalar field in the radial direction. In Ref. \cite{ESCback}, the holographic dual of the excited state superconductor with the full back-reaction is also constructed numerically. It is found that there are additional peaks in the imaginary and real parts of the electrical conductivity and the number of peaks corresponds to the number of nodes of the $n$-th excited state.

As alluded to above, the studies of Ref. \cite{ESC} and \cite{ESCback} focused exclusively on the linear response in the holographic model of the excited state superconductors (see also \cite{Qiao:2020fiv} where an analytical study is performed). In condensed matter physics, especially for the mesoscopic nanomaterials, under the thermal fluctuation, the system may transit to metastable states and remain in these states for a long time due to the complicated free energy surface \cite{PSBD}. On the other hand, in the modern experiments of the condensed matter physics, it is possible to drive a system into a far-from-equilibrium state and measure how the system settles down towards the equilibrium state\cite{PSSV}. Therefore, it is very important to investigate such a relaxation process in holographic superconductors. This may provide us with useful information about the non-equilibrium superconductivity that can be compared to the experiments in the condensed matter physics \cite{NSbook,LS}.

With this in mind, we shall work in the micro-canonical ensemble and investigate the non-equilibrium condensation process triggered by the perturbation on the normal state in the holographic superconductor to see the role played by the aforementioned excited states. As such, we are required to evolve the full non-linear gravitational dynamics of the Einstein-Maxwell theory coupled with the complex scalar field in asymptotic AdS space with the initial perturbations somewhat different from those used in the previous studies on the non-equilibrium condensation process or the quench process of the ground states of holographic superconductors \cite{
BGSSW,LTZ,GGZZ,GZZ,BMN,SCZ,ZengZhang,VS,LXZZ,GKLTZ,KSN,XBMAni,YYB,ranprd,pwave,LiMG}, or the holographic isotropization of non-Abelian plasma \cite{Heller:2012km,Heller:2013oxa}. For example, in \cite{KSN,XBMAni,YYB,ranprd,pwave,LiMG}, the Gaussian wave packets are used as the initial perturbations of the scalar fields to probe the dynamics of the holographic superconductors. The Gaussian type of the initial data is also used to study the nonlinear evolution and the final state of the spherical symmetric RNAdS black hole due to the superradiant instability \cite{BGLPRL}. While in this paper, the initial perturbations we take to evolve the system are the eigenfunctions of the linear perturbation equation of the scalar field in the background of the planar RNAdS black hole or their linear superposition. This type of the initial perturbations was originally proposed in \cite{BGLR} to study the dynamical formation of the excited states of spherically symmetric hairy black hole in AdS space. In this way, we can observe the dynamical formation of the excited states as the intermediate states during the relaxation from the normal state to the ground state. We also provide a global and physical picture of the evolution dynamics of the black hole and the corresponding superconducting phase transition from the funneled landscape view, quantifying the weights of the states and characterizing the transitions and cascades towards the ground state.

The paper is structured as follows. In Section \ref{setup}, we will present the setup of the holographic superconductor model and describe the numerical method to solve the gravitational dynamical equations. In Section \ref{numerical}, after discussing the initial perturbations used in the time evolution process, we present the numerical results of the non-equilibrium dynamical process of the holographic superconductors with the excited states as the intermediate states, including the time dependence of the scalar field, the superconducting order parameter, the areas of the event and apparent horizons. We conclude our paper with some discussions in Section \ref{end}.

\section{Holographic setup}\label{setup}

We consider the holographic $s$-wave superconductor described by the Einstein-Maxwell-complex scalar field theory with a negative cosmological constant. The action in the four dimensions is given by \cite{3HJHEP}
 \begin{equation} \label{action}
S=\int d^4 x\sqrt{-g}\left[R+\frac{6}{l^2}-\frac{1}{4}F_{\mu\nu}F^{\mu\nu}-|\nabla\Psi-iqA|^2 -m^2|\Psi|^{2}\right],
\end{equation}
where $F_{\mu\nu}=\nabla_{\mu}A_{\nu}-\nabla_{\nu}A_{\mu}$ is the $U(1)$ gauge field strength,
$l$ is the curvature radius of the AdS space, and $m$ and $q$ represent the mass and the charge of the complex scalar field $\Psi$ respectively. By varying with respect to the metric $g_{\mu\nu}$, the complex scalar field $\Psi$, and the $U(1)$ gauge potential $A_{\mu}$, one can derive the Einstein equation, the Klein-Gordon equation, and the Maxwell equation as follows
\begin{eqnarray}
&&R_{\mu\nu}-\frac{1}{2}g_{\mu\nu}R-\frac{3}{l^2}g_{\mu\nu}
=T^{EM}_{\mu\nu}+T^{\Psi}_{\mu\nu}\;,\label{equation0}\\
&&(\nabla_\mu-iqA_\mu)(\nabla^\mu-iqA^\mu) \Psi -m^2\Psi=0\;,\label{equation1} \\
&&\nabla_{\mu} F^{\mu\nu}=iq[ \Psi^{*}(\nabla^\nu-iqA^\nu)\Psi- \Psi(\nabla^\nu+iqA^\nu)\Psi^{*}]\;,\label{equation2}
\end{eqnarray}
where the stress-energy tensors of the matter sector are given by
\begin{eqnarray}
T^{EM}_{\mu\nu}&=&\frac{1}{2}\left(g^{\sigma\rho}F_{\mu\sigma}F_{\nu\rho}
-\frac{1}{4}g_{\mu\nu}F^{\rho\sigma}F_{\rho\sigma}\right)\;,\\
T^{\Psi}_{\mu\nu}&=&
\frac{1}{2}[(\nabla_{\mu}\Psi-iqA_\mu\Psi)(\nabla_{\nu}\Psi^{*}+iqA_\nu\Psi^{*})
+c.c.]\nonumber\\
 &&-\frac{1}{2}g_{\mu\nu}|\nabla\Psi-iqA\Psi|^2 -\frac{1}{2}g_{\mu\nu}m^2|\Psi|^{2}\;.
\end{eqnarray}

It is well known that the characteristic formulation has been successfully applied in studying a variety of gravitational dynamics problems in asymptotically AdS spacetime \cite{CY}.
Following this approach, we adopt the following metric ansatz in the ingoing Eddington-Finkelstein coordinates
\begin{equation}\label{metric}
ds^2=-\frac{1}{z^2}\left[F(v,z)dv^2+2dvdz\right]+\Phi(v,z)^2(dx^2+dy^2).
\end{equation}
We work in a axial gauge where the ansatz of the gauge field and the scalar field can be written as
\begin{equation}\label{matter}
A=\alpha(v,z)dt,\;\;\;\Psi=\psi(v,z).
\end{equation}
It should be noted that diffeomorphism and $U(1)$ gauge symmetries are not completely fixed by the above ansatz. These residual gauge symmetries are given by
\begin{equation}\label{residualgauge}
\frac{1}{z}\rightarrow\frac{1}{z}+g(v),~\;\alpha\rightarrow\alpha+\partial_v\theta(v),~\;
\psi\rightarrow\psi e^{iq\theta(v)}\;,
\end{equation}
which can be further fixed by taking into account of the boundary conditions later on. Hereafter, without loss of generality, we set the value of the curvature radius of the AdS space $l$ to be one.

By substituting the ansatz of the metric and the matter fields into the Einstein equation, the Klein-Gordon equation, and the Maxwell equation, one can derive the equations of motion as follows
\begin{eqnarray}
&&\Phi''+\frac{2}{z}\Phi'+\frac{1}{2}\Phi|\psi'|^2=0,\label{eq1}\\
&&\alpha''+2\left(\frac{1}{z}+\frac{\Phi'}{\Phi}\right)\alpha'
+\frac{iq}{z^2}(\psi\psi'^{*}-\psi^{*}\psi')=0,\\
&&(D\Phi)'+\frac{\Phi'}{\Phi}D\Phi-\frac{\Phi}{8z^2}(z^4\alpha'^2
+2m^2|\psi|^{2}-12
)=0,\\
&&(D\psi)'+\frac{\Phi'}{\Phi}D\psi+\frac{D\Phi}{\Phi}\psi'+
\frac{1}{2z^2}(iq z^2\alpha'+m^2)\psi=0,\label{psieq}\\
&&\left(z^2\left(z^{-2} F\right)'\right)'-z^2\alpha'^2+4\Phi^{-2}\Phi'D\Phi
-(\psi'^{*}D\psi+\psi'D\psi^{*})
=0,\label{eq5}
\end{eqnarray}
and
\begin{eqnarray}
&&2z^2(D\alpha)'+(4z^2\Phi^{-1}D\Phi
+z^2F'-2zF)\alpha'+2iq(\psi^{*}D\psi-\psi D\psi^{*})=0,\label{constr1} \\
&&zD^2\Phi-FD\Phi+\frac{1}{2}z(\Phi D\psi D\psi^{*}+D\Phi F')+\frac{F^2}{8}(4\Phi'+z\Phi|\psi'|^2+2z\Phi'')=0,\label{constr2}
\end{eqnarray}
where the prime denotes the derivative with respect to the radial coordinate $z$. The derivative operator $D$ for the real field is defined as $D=\partial_v-\frac{1}{2}F \partial_z$, which is the derivative along the radial outgoing null geodesics, while the derivative operator $D$ for the complex scalar filed $\psi$ is defined as
\begin{eqnarray}
D\psi=\partial_v\psi-\frac{1}{2}F\partial_z\psi-iq\alpha\psi.
\end{eqnarray}

We briefly describe the numerical procedure to solve the equations of motion as follows.
The equations (\ref{eq1})-(\ref{eq5}) are regarded as the evolution equations
and the equations (\ref{constr1})-(\ref{constr2}) are regarded as the constraint equations.
At the initial moment, with the initial configuration for the scalar field $\psi$, we can successively solve the evolution equations (\ref{eq1})-(\ref{eq5}) to get the functions $\Phi$, $\alpha$, $D\Phi$, $D\psi$, and $F$ by using the Chebyshev pseudo-spectral method. The time derivative of the scalar field $\partial_v\psi$ at this time step can be extracted from the definition of $D\psi$. We can evolve the scalar field configuration to the next time step by using the forth order Runge-Kutta method. Repeating the above procedure, we can obtain the full time dependent solution of the system.
The constraint equations (\ref{constr1})-(\ref{constr2}) can be used to check the accuracy of the numerical solution.

The asymptotic behaviors of the fields provide the boundary conditions which are used to solve the evolution equations. We consider $m^2=-2$ for simplicity. By solving the equations of motion (\ref{eq1})-(\ref{constr2}) near the AdS boundary $(z=0)$ order by order, we obtain the asymptotic expansions as follows
\begin{equation}
\begin{split}
&F(v,z)=1 -2M z^3 +\left(\frac{Q^2}{4}-\frac{|\psi_2|^2}{3}\right)z^4+ \cdots\ ,\\
&\Phi(v,z)=\frac{1}{z} +\Phi_{0} - \frac{|\psi_2|^2}{6}z^3 + \cdots\ ,\\
&\alpha(v,z)=\alpha_0 +Qz
+\frac{iq}{12}\left(\psi_2^*\dot{\psi}_2-\psi_2\dot{\psi}_2^*\right)z^4 +\cdots\ ,\\
&\psi(v,z)=\psi_2z^2 + \dot{\psi}_2 z^3+\cdots\ ,
\label{asympt}
\end{split}
\end{equation}
where the dot means $\frac{d}{dv}$. It should be noted that we have taken the first order expansion coefficient $\psi_1$ of the scalar field as the source and set $\psi_1=0$. Therefore, according to holographic renormalization \cite{KS}, the second order expansion coefficient $\psi_2$ should be identified as the vacuum expectation value of the operator $\hat{O}$ in the dual field theory on the AdS boundary.

Note that there are two parameters $M$ and $Q$ in the asymptotic expansions, which represent the ADM mass and the charge of the solution, respectively. As we shall work in the micro-canonical ensemble, these two parameters will be fixed during the whole evolution. The functions $\Phi_{0}(v)$ and $\alpha_0(v)$ represent the residual gauge freedoms presented in Eq. (\ref{residualgauge}). As discussed in \cite{CY}, the freedom of $\Phi_{0}(v)$ can be used to fix the position of the apparent horizon at a constant radial coordinate during the time evolution. However, we do not explore the dynamics of $\Phi_{0}(v)$. It is convenient to make further gauge choice that $\Phi_{0}(v)=\alpha_0(v)=0$.

After fixing the residual gauge freedoms, the function $\psi_2(v)$ is the only unknown function which shall be determined by the numerical solution. It can be observed that the fields $\Phi$ and $D\Phi$ are divergent at the AdS boundary. Therefore, following the general strategy \cite{KSN,CY}, the field redefinitions prior to numerical calculation should be introduced to remove the divergence of the fields at the AdS boundary. The equations of motion for the redefined fields can be derived from Eq. (\ref{eq1})-(\ref{constr2}) easily and the corresponding boundary conditions at the AdS boundary can be obtained from the asymptotic expansions (\ref{asympt}). With these in hand, we can evolve the whole system from arbitrary initial data and extract the time dependent behaviors of the scalar field, the superconducting order parameter, and the geometry of the background hairy black hole.

\section{Numerical results}\label{numerical}

Until now, the initial conditions of the time evolutions are not specified. Before presenting the numerical results, let us discuss the initial conditions we shall use in the time evolutions of the system.

\subsection{Initial conditions}

Firstly, we discuss the initial background spacetime. The initial background spacetime is taken as the bald RNAdS black hole with planar symmetry. When the scalar field is null, the solution to the equations of motion is the static RNAdS black hole solution with
\begin{eqnarray}\label{RNAdS}
F=1-2Mz^3+\frac{1}{4}Q^2 z^4,~~\Phi=\frac{1}{z},~~\alpha=Qz,~~\psi=0.
\end{eqnarray}
This solution can be regarded as the normal phase of a holographic superconductor. We take this bald RNAdS black hole as the initial spacetime background. Notice that the equations of motions (\ref{eq1})-(\ref{constr2}) have the following scaling symmetry
\begin{align}
 &(v,z,x,y)\rightarrow(kv,kz,kx,ky)\ ,\\
 &F \rightarrow F\ ,\quad \Phi \rightarrow \Phi/k\ ,\quad \alpha \rightarrow \alpha/k\ ,\quad \psi \rightarrow \psi\ ,\\
 &M \rightarrow M/k^3\ ,\quad Q \rightarrow Q/k^2\ ,\quad T \rightarrow T/k\, .
\label{scaling}
\end{align}
Using this scaling symmetry, we can set the horizon radius of the initial RNAdS black hole to unity without the loss of the generality. The mass and the temperature of the black hole are then given by
\begin{equation}
M=\frac{1}{2}\left(1+\frac{1}{4}Q^2\right)\;,\;\;T=\frac{12-Q^2}{16\pi}\;,
\end{equation}
which implies that the initial spacetime background is only determined by the black hole charge $Q$.

Then, we discuss the initial perturbation of the scalar field. The charged scalar field in the background of the planar RNAdS black hole is unstable due to the violation of the BF bound near the horizon. As mentioned in the introduction, we use the eigenfunctions of the linearized equation of motion of the scalar field in the background of the planar RNAdS black hole or their linear superposition as the initial perturbations of the scalar field. For this purpose, we should study the quasi-normal modes of the scalar field perturbations in the background of the planar RNAdS black holes. The quasi-normal modes of the planar RNAdS black holes are studied in \cite{Maeda:2006by,MMZ,AKL,Janiszewski:2015ura,Ammon:2017ded,Abbasi:2020ykq}.

The linearized equation of motion of the scalar field, which can be directly reduced from Eq.(\ref{psieq}) by replacing the metric functions via those of the planar RNAdS black hole, is given explicitly by
\begin{eqnarray}\label{linearizedEq}
\psi''(z)+\left(\frac{F'(z)+2i\left(\omega+q Q z\right)}{F(z)}-\frac{2}{z}\right)\psi'(z)
-\frac{\left(i q Q z^2 +2 i \omega z +m^2\right)}{z^2 F(z)}\psi (z)=0\;,
\end{eqnarray}
where the metric function $F(z)$ is given in Eq.(\ref{RNAdS}). As discussed above, the location of the horizon of the initial RNAdS black hole is set to be unity. Thus there are only two parameters, i.e. the black hole charge $Q$ and the scalar field charge $q$, left to change freely.

The quasi-normal modes can then be computed directly by treating the linearized equation of motion as an eigenvalue problem. The linearized equation of motion (\ref{linearizedEq}) can be casted into the generalized eigenvalue equation in the matrix form by using the Chebyshev pseudo-spectral method \cite{RanSuper}. One can refer to Ref. \cite{AJ} for a detailed discussion of the method for computing quasi-normal modes in terms of the pseudo-spectral method. By solving the linearized equation under the boundary condition that $\psi$ is ingoing at the horizon and zero at the AdS boundary, one can obtain the quasi-normal modes of the scalar field that are characterized by the overtone number $n$. Then, after substituting the quasi-normal modes into the linearized equation of motion of the scalar field, the resulting equation can be solved to obtain the corresponding eigenfunctions $\psi_{(n)}$.
By varying the two parameters $Q$ and $q$, one can obtain more than one unstable modes in the spectrum of the linearized equation of motion. In the following, we will take the eigenfunctions of the unstable modes or their linear superposition as the initial perturbations in the nonlinear evolutions.

\subsection{Case I: The spectrum of scalar perturbation has two unstable modes }

\begin{table}
\caption{Quasinormal modes of scalar field in the planar RNAdS black hole for $Q=2$ and $q=6.5$. There are two unstable modes in the spectrum.}
\centering
\begin{tabular}{lcccc}
  \\
  \hline
  \hline
  \;\;\;$n$\;\;\; & $\omega_n$\\
  \hline
  \;\;\;0\;\;\; & \;\;\;-7.7538321384+0.2558913563i\;\;\; \\
  \hline
  \;\;\;1\;\;\; & \;\;\;-11.2264046848+0.1381951341i\;\;\; \\
  \hline
  \;\;\;2\;\;\; & \;\;\;-13.0249987251-0.008106445792i\;\;\; \\
  \hline
  \hline
\end{tabular}\label{QNMcaseI}
\end{table}

\begin{figure}
\centering
  \includegraphics[width=10cm]{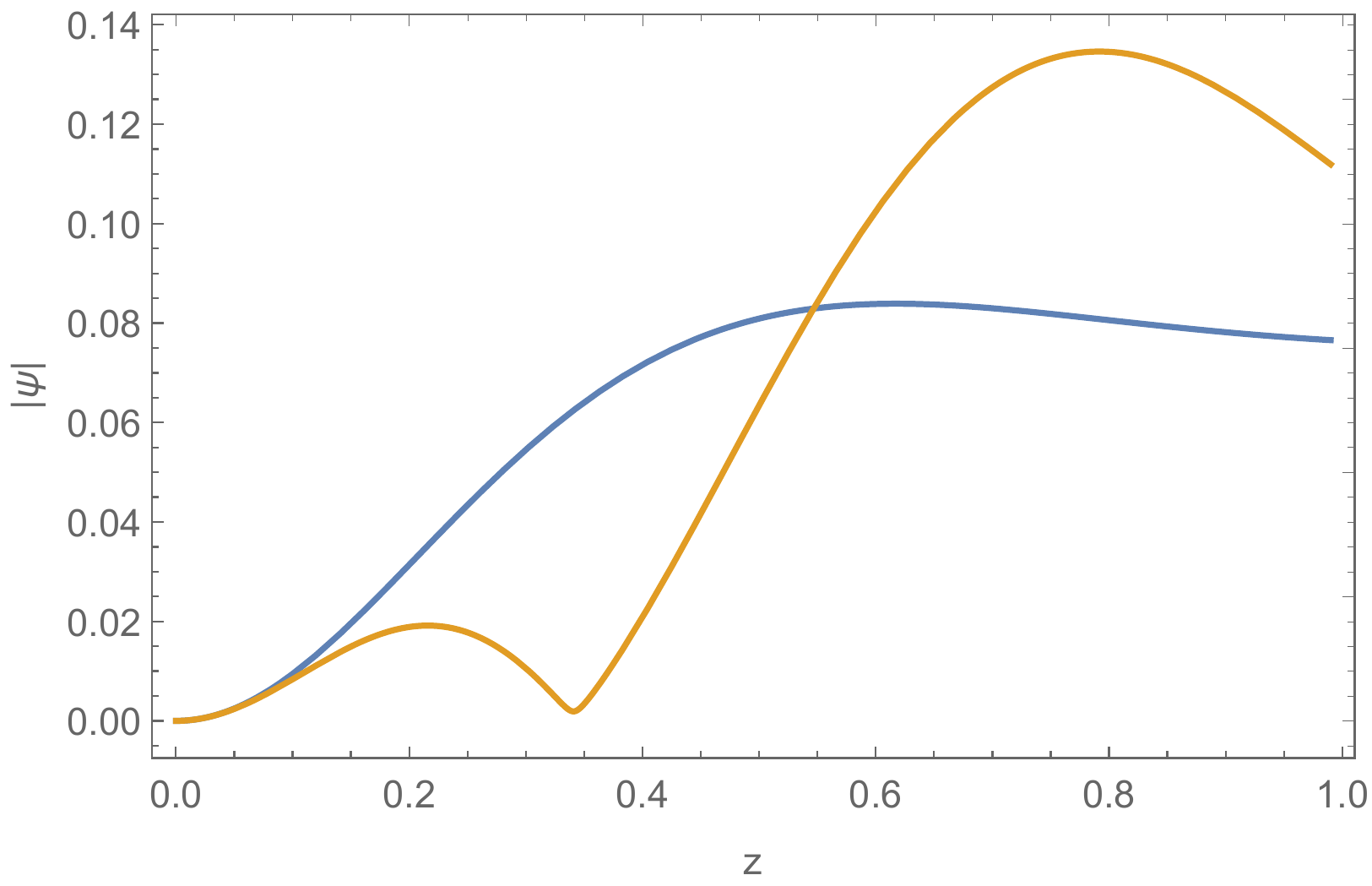}\\
  \caption{The eigenfunction $\psi_{(0)}$ (blue line) and $\psi_{(1)}$ (yellow line) of the scalar field for $Q=2$ and $q=6.5$. The overtone number corresponds to the node number.}\label{EigenFuncI}
\end{figure}

Firstly, we consider the case of $Q=2$ and $q=6.5$, in which the spectrum of the linearized equation of motion for the scalar field has only two unstable modes.
The quasinormal modes are listed in the Table \ref{QNMcaseI}. It is shown that the growth rate $\textrm{Im}[\omega_n]$ decreases as the overtone number $n$ grows, while the oscillation frequency $\textrm{Re}[\omega_n]$ increases. The two modes with the overtone number $n=0, 1$ are unstable. The corresponding eigenfunctions $\psi_{(0)}$ and $\psi_{(1)}$ are plotted in Fig. \ref{EigenFuncI}. It is observed that the node number can be used to characterize the eigenfunction.

If we choose the initial perturbation of the scalar field as the Gaussian wave pocket or the first eigenfunction $\psi_{(0)}$ of the linearized equation of motion, the system will evolve to the ground state directly. This kind of dynamical process has been investigated previously in \cite{KSN,XBMAni,YYB,ranprd,pwave,LiMG}, which can be regarded as the nonequilibrium condensation process of the ground state of holographic superconductor.

In order to study the non-equilibrium condensation process with the $n=1$ excited state formed as the intermediate state, we choose the second eigenfunction $\psi_{(1)}$ of the linearized equation of motion for the scalar field as the initial perturbation in the nonlinear evolution \cite{BGLR}. The relevant numerical results are detailed as follows.

\begin{figure}
\centering
  \includegraphics[width=10cm]{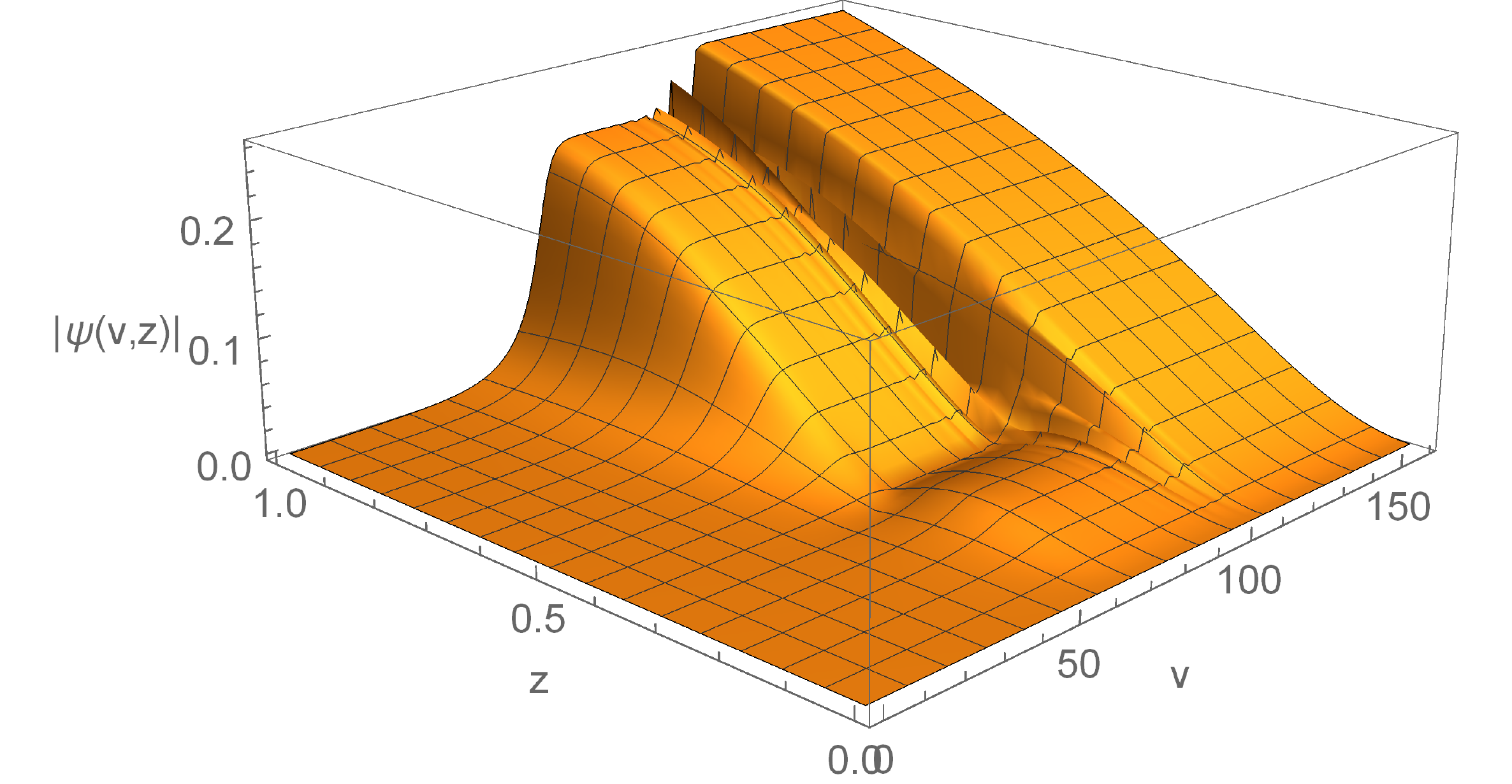}\\
  \caption{The dynamics of scalar field for $Q=2$ and $q=6.5$. The second eigenfunction of the linearized equation of motion is taken to be the initial condition for the scalar field. }\label{psicaseI}
\end{figure}

The dynamics of the scalar field is presented in Fig. \ref{psicaseI}. As one can see, in accord with the fact $\psi_{(1)}$ is a linear unstable mode, the amplitude of the scalar field is initially amplified exponentially to a metastable intermediate state with two visible crests along the radial direction, which corresponds to the excited hairy black hole with the overtone number $n=1$. This dynamical process can be identified as the nonequilibrium condensation process of the excited state of holographic superconductor. Note that this dynamical process from the initial RNAdS black hole to the intermediate state is non-linear rather than linear. So generically the $n=0$ mode will be generated. Due to its higher increasing rate than $n=1$ mode, it will eventually overshadow the $n=1$ mode and drive the scalar field away from the excited state to the ground state of the holographic superconductor, which is confirmed by the temporal profile demonstrated in Fig. \ref{psicaseI}.
However, during the whole dynamical process, the holographic superconductor can remain in the excited state for a considerable time period, which provides us with a good manner to prepare the excited holographic superconductor.

\begin{figure}
\centering
  \includegraphics[width=10cm]{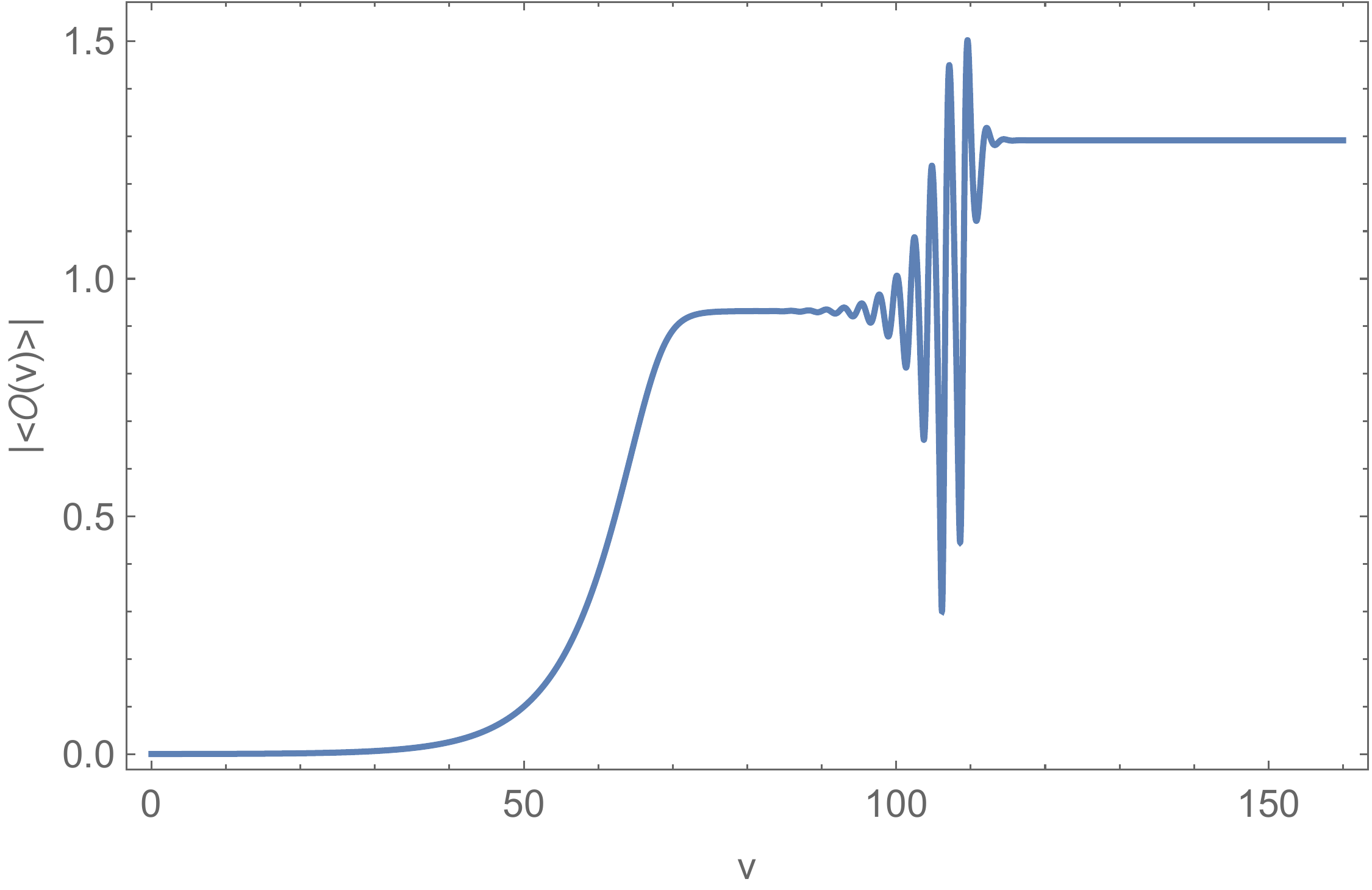}\\
  \caption{The time evolution of superconducting order parameter $\langle \mathcal{O}\rangle$ for $Q=2$ and $q=6.5$.}\label{OcaseI}
\end{figure}

The time evolution of the superconducting order parameter $|\langle \mathcal{O}\rangle|$, which is proportional to $|\psi_2(v)|$, is presented in Fig.\ref{OcaseI}. Obviously, there are two plateaus in the temporal profile of the superconducting order parameter. The superconducting order parameter evolves smoothly to the nonzero value at the first plateau and then reaches the final value in an oscillating way. This strongly manifests that during the evolution the superconductor on the boundary reaches the excited state, signalled by the first plateau. However, because the intermediate state is metastable, the boundary system eventually evolves to the final ground state, corresponding the second plateau. As discussed above, during the dynamical evolution, the generation of $n=1$ mode is inevitable. Therefore, the oscillating behavior from the excited state to the ground state can be understood roughly as the interference between the $n=0$ and $n=1$ modes, i.e.,
\begin{eqnarray}
|\langle \mathcal{O}\rangle|&\sim& |a e^{-i\omega_0 v}+b e^{-i\omega_1 v}|\nonumber\\
&\sim& \left[A(v)+B(v)\cos\left(|\textrm{Re}[\omega_0-\omega_1]|v \right)\right]^{1/2}\;,
\end{eqnarray}
whereby the oscillation frequency of the transition process is then determined by the difference of the real parts of the unstable modes.

\begin{figure}
\centering
  \includegraphics[width=10cm]{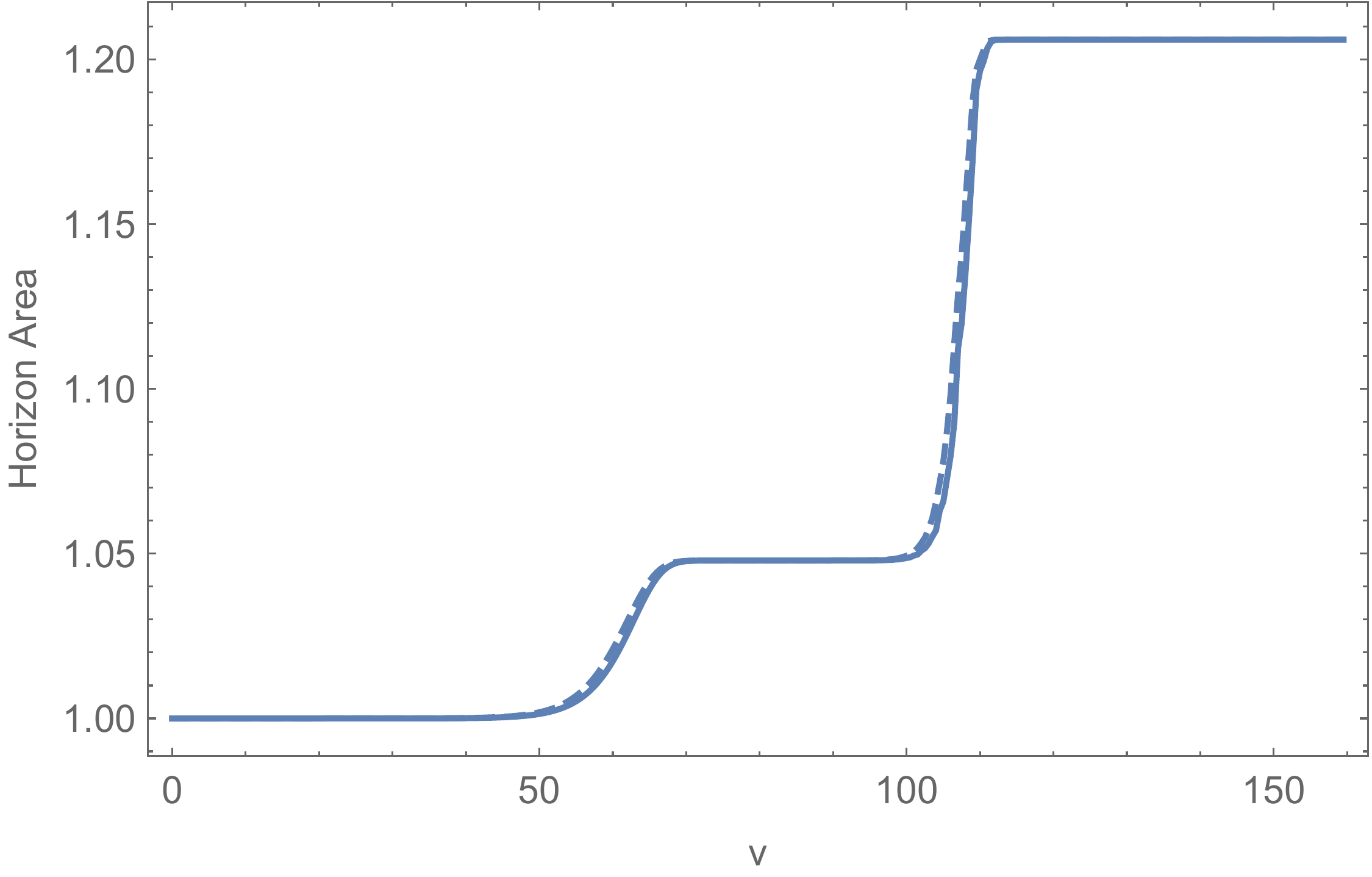}\\
  \caption{The time evolutions of the areas of event horizon (dashed line) and apparent horizon (solid line) for $Q=2$ and $q=6.5$. The excited state of holographic superconductor occurs at the first plateau.}\label{horizoncaseI}
\end{figure}

The areas of the event horizon and the apparent horizon are presented in Fig. \ref{horizoncaseI}. This figure shows the evolution of the bulk black hole geometry, which also implies that the system reaches the intermediate excited state during the time evolution. In contrast to the behaviors of the scalar field and the superconducting order parameter, the transition processes of the horizons from the intermediate excited state to the final stationary state are smooth. Another observation is that the areas of the horizons monotonously increases with the time, which is consistent with the second law of black hole thermodynamics. We can also see that the event horizon always has a larger area than that of the apparent horizon during the transition processes from the bald RNAdS black hole to the intermediate state and from the intermediate state to the final stationary state. When the system is in the metastable intermediate state or in the final stationary state, the apparent horizon and the event horizon coincide, which implies the formation of a hairy black hole at the excited state and at the ground state.

\begin{figure}
\centering
  \includegraphics[width=10cm]{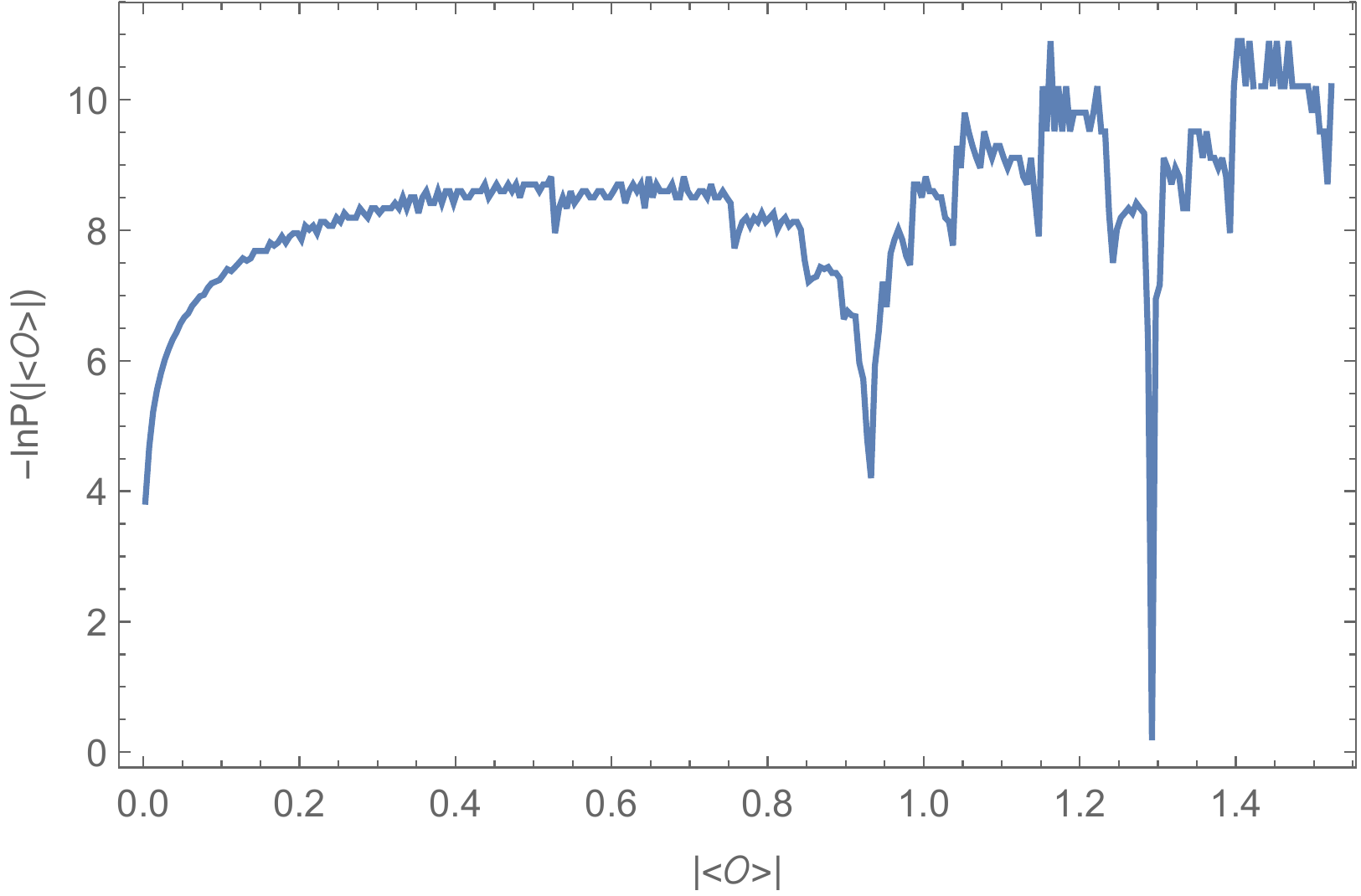}\\
  \caption{Probability distribution of the superconducting order parameter in the whole evolution process for $Q=2$ and $q=6.5$. Typically, the order parameter is equally discretized with the interval $\delta=0.005$. }\label{LnPcaseI}
\end{figure}

To get a physical picture and better understanding of the evolution process, we analyze the time trace of the superconducting order parameter $|\langle \mathcal{O}\rangle|$. We collect the corresponding trajectory statistics and therefore quantify the probability distribution $P(|\langle \mathcal{O}\rangle|)$ of the order parameter. The value of the order parameter represents the state of the system. The probability $P(|\langle \mathcal{O}\rangle|)$ gives the weight of each state of the system. If we consider the microcanonical ensemble with the fixed mass and charge, the probability distribution $P(|\langle \mathcal{O}\rangle|)$ can be related to the entropy of the system $S\sim\ln P$. On the other hand, in the canonical ensemble at certain fixed temperature, the probability distribution $P(|\langle \mathcal{O}\rangle|)$ can be related to the effective potential or free energy $F(|\langle \mathcal{O}\rangle|)/kT\sim -\ln P(|\langle \mathcal{O}\rangle|)$. We displayed this effective potential or free energy landscape in Fig. \ref{LnPcaseI} with respect to the order parameter $|\langle \mathcal{O}\rangle|$. Note that the landscape quantifies the importance or the weight of each state. This leads to a global picture and characterization of the states of the system.

From Fig. \ref{LnPcaseI}, the effective potential has three wells at $|\langle \mathcal{O}\rangle|=0$, $0.92$, and $1.3$, which corresponds to the normal state, the excited state, and the ground state of the superconductor on the bondary and the bald RNAdS black hole, the excited state back hole, and the ground state black hole in the bulk. The depth of the potential well quantifies the stability of the state in the evolution process. It is clear that the ground state black hole is more stable while the excited state black hole is less stable or metastable. The corresponding superconducting state on the boundary is more stable while the excited state is less stable or metastable. If the evolution time approaches to infinity, the probability that the system stays at the ground state black hole or the corresponding superconducting ground state approaches to one. The overall shape of the landscape is a funnel towards the ground state either for hairy black hole in the bulk or superconducting state at the bondary with certain bumpiness or roughness along the way (intermediate basins), when the initial temperature is below the critical transition temperature $T_c$. In fact, the funnel provides a guiding force of the evolution towards the ground state of the hairy black hole or the corresponding superconducting state. Here we can see a clear physical picture of the evolution process for the black hole in the bulk and the corresponding phase transition dynamics of the superconductor on the boundary. Starting from the normal state or the bald RNAdS black hole state basin, the system needs to pass through a potential barrier to reach the excited state black hole. Since the intermediate transient state is metastable, the evolution continues over another potential barrier to the superconducting ground state or the ground state black hole. Here the depth of the basin gives the information of the time duration staying in the corresponding state while the barrier height provides the measure of the degrees of the difficulties of the transition or switching from one state to another.
Therefore this funneled effective landscape with roughness not only can identify the important states (depth) but also provide global insight on the chance and speed of the transitions and cascade across barriers towards the ground state.

\subsection{Case II: The spectrum of scalar perturbation has multiple unstable modes}

\begin{table}
\caption{Quasinormal modes of scalar perturbation for $Q=3.4$ and $q=6$. There are multiple unstable modes in the spectrum.}
\centering
\begin{tabular}{lcccc}
\\
  \hline
  \hline
  \;\;\;$n$\;\;\; & $\omega_n$\\
  \hline
  \;\;\;0\;\;\; & \;\;\;-9.7248526672+0.3244771629i\;\;\; \\
  \hline
  \;\;\;1\;\;\; & \;\;\;-14.1765978642+0.1863067015i\;\;\; \\
  \hline
  \;\;\;2\;\;\; & \;\;\;-16.8693979997+0.1193051324i\;\;\; \\
  \hline
  \;\;\;3\;\;\; & \;\;\;-18.5543016416+0.07427889118i\;\;\; \\
  \hline
  \;\;\;4\;\;\; & \;\;\;-19.5549744811+0.04250028752i\;\;\; \\
  \hline
  \;\;\;5\;\;\; & \;\;\;-20.0951387943+0.02157862614i\;\;\; \\
  \hline
  \;\;\;6\;\;\; & \;\;\;-20.3478427333+0.008659087290i\;\;\; \\
  \hline
  \;\;\;7\;\;\; & \;\;\;-20.4267071604-0.02354199123i\;\;\; \\
  \hline
  \hline
\end{tabular}\label{QNMcaseII}
\end{table}

\begin{figure}
\centering
  \includegraphics[width=10cm]{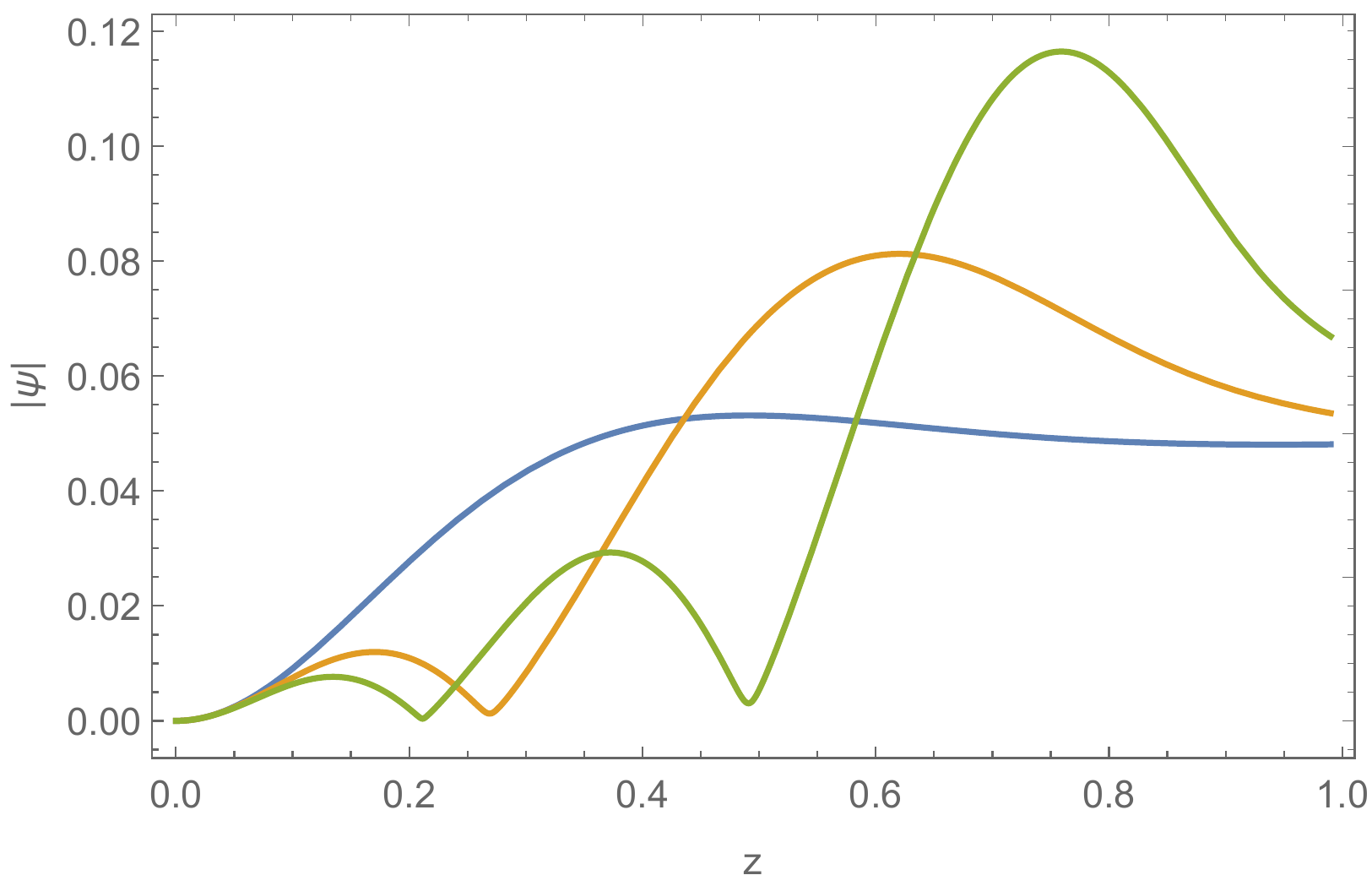}\\
  \caption{The first three eigenfunctions $\psi_{(1)}$ (blue line), $\psi_{(2)}$ (brown line), and $\psi_{(3)}$ (green line) of the linearized equation of motion of the scalar field for $Q=3.4$ and $q=6$. }\label{EigenFuncII}
\end{figure}

In this subsection, we consider the case  of $Q=3.4$ and $q=6$, which is more complicated in the sense that the spectrum of the scalar perturbation has multiple unstable modes. The corresponding quasi-normal modes of the linearized scalar field are listed in the Table \ref{QNMcaseII}. It is also shown that the growth rate $\textrm{Im}[\omega]$ decreases and the oscillation frequency $\textrm{Re}[\omega]$ increases as the overtone number $n$ grows. The first three eigenfunctions are plotted in Fig. \ref{EigenFuncII}. It is also observed that the node number can be used to characterize the eigenfunction.

We consider two types of special initial data. The first type of initial data consists of a single overtone mode $\psi_{(n)}$ with $n\neq 0$. As shown in the last subsection, we expect that the system initially evolves into an excited hairy black hole with the overtone number $n$, corresponding to the excited state of the holographic superconductor, and then decays to the hairy black hole with overtone number $n=0$, corresponding to the ground state superconductor. The behavior for the single mode initial data should be qualitatively similar to the case considered in the last subsection.

The second type of the initial data is a mixture of the two modes, i.e., the linear superposition of the two eigenfunction $\psi_{(1)}$ and $\psi_{(2)}$:
\begin{eqnarray}
\psi_{mix}=a_{(1)}\psi_{(1)}+a_{(2)}\psi_{(2)}\;
\end{eqnarray}
with $a_{(1)}/a_{(2)}=1/999$. With this type of initial perturbation, we want to see the dynamical transition process between the excited states of holographic superconductor. Namely, we hope to achieve a cascade, where initially the $n=2$ excited black hole forms, which then decays to $n=1$, and finally to $n=0$.

\begin{figure}
\centering
  \includegraphics[width=8cm]{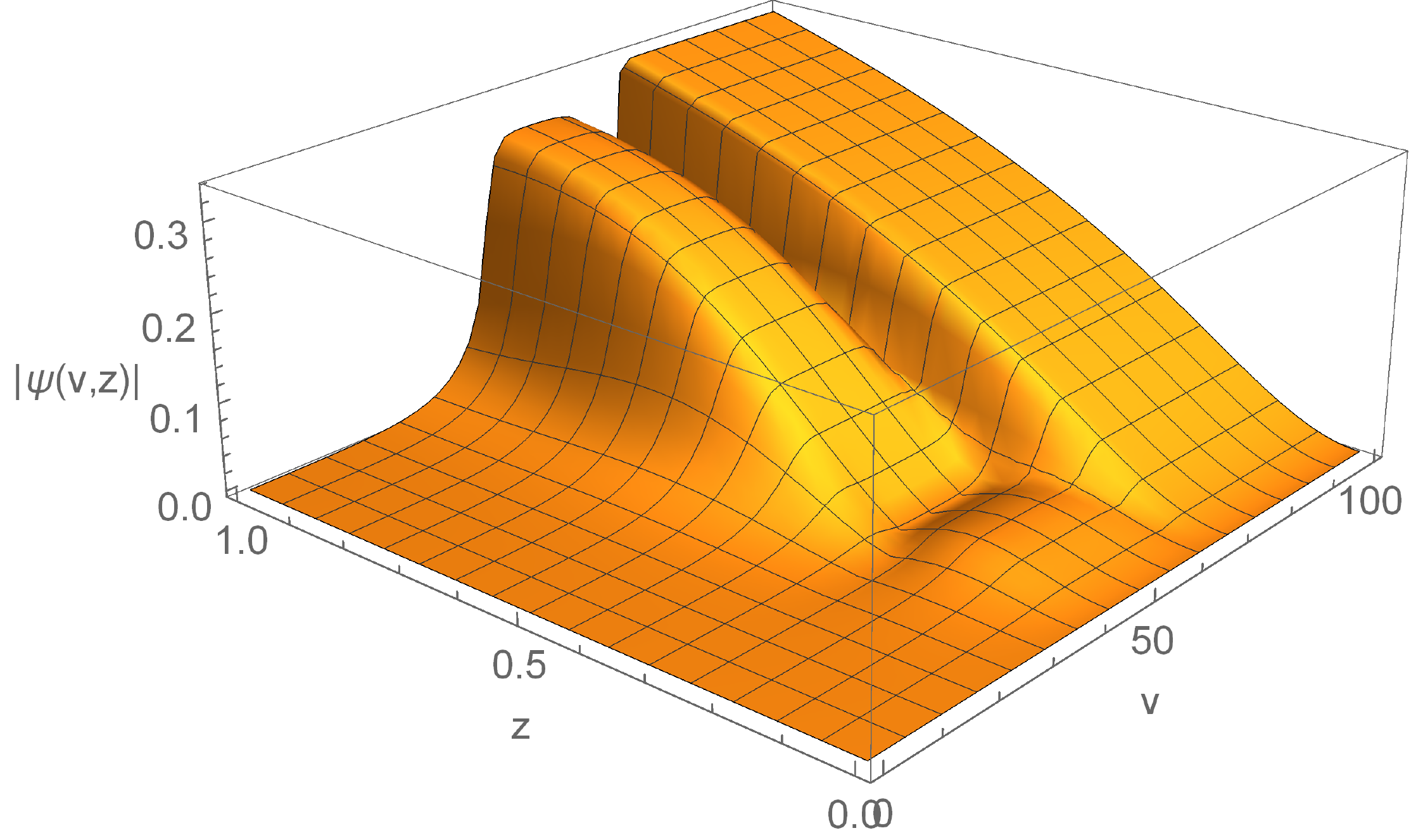}\\
  \includegraphics[width=8cm]{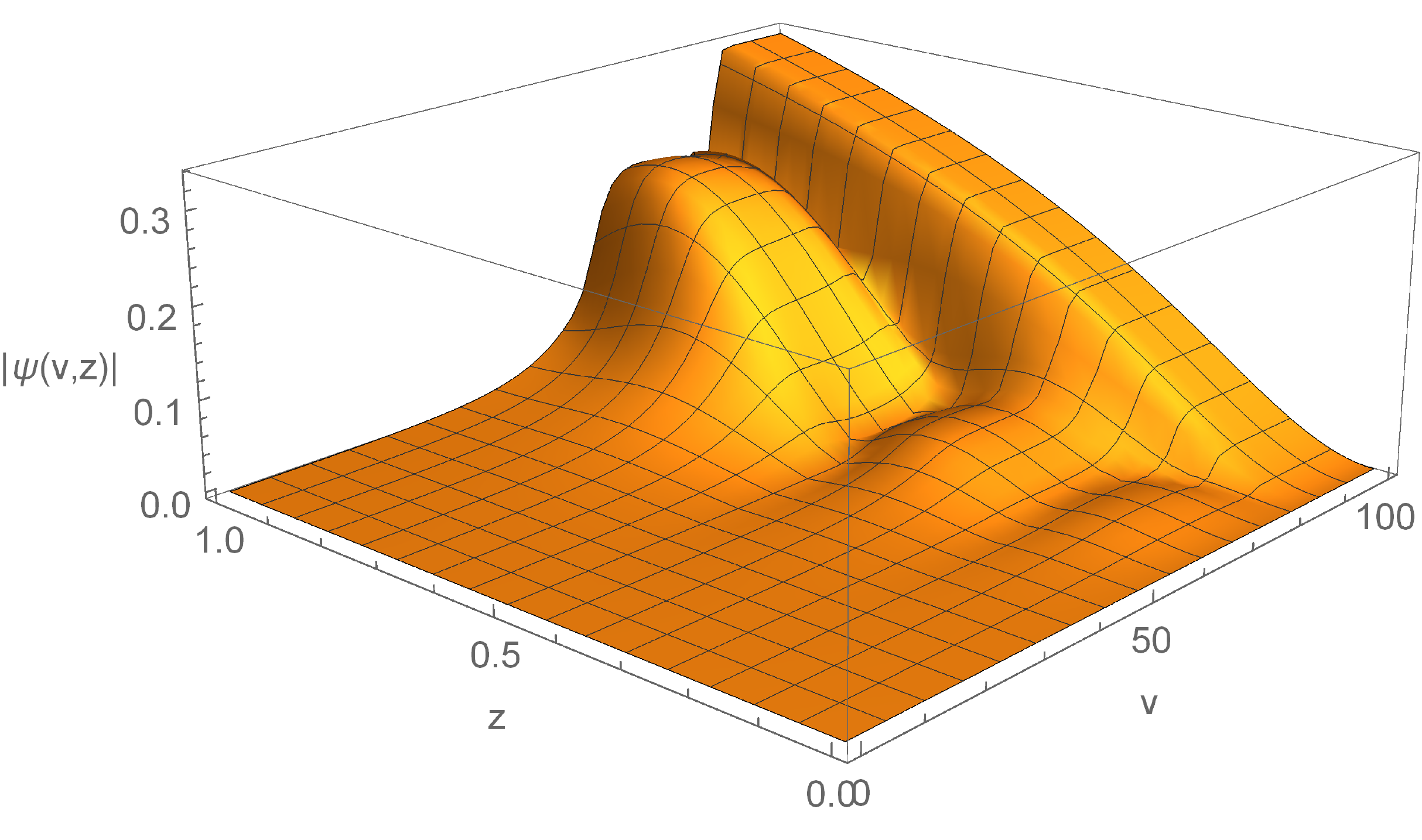}\\
  \includegraphics[width=8cm]{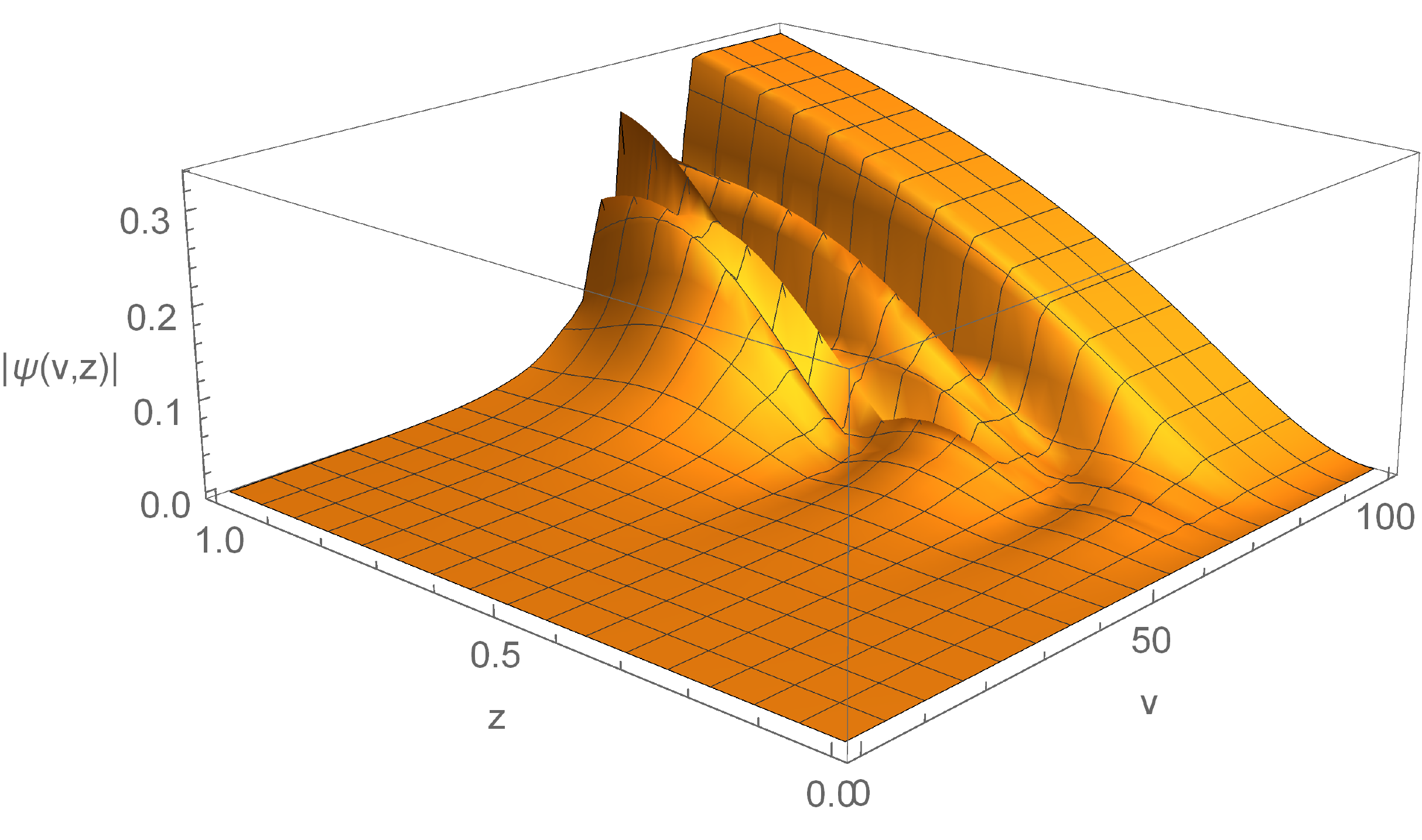}\\
  \caption{The dynamics of the scalar field for $Q=3.4$ and $q=6$. In the three panels, the initial perturbation are taken as the second eigenfunction $\psi_{(1)}$, the third eigenfunction $\psi_{(2)}$, and the superposition $\psi_{mix}$ of the second and the third eigenfunctions. }\label{psidynamicscaseII}
\end{figure}

In Fig. \ref{psidynamicscaseII}, we show that the dynamics of the scalar field for $Q=3.4$ and $q=6$ with three kinds of initial perturbations. In the three panels, the initial perturbations are taken as the second eigenfunction $\psi_{(1)}$, the third eigenfunction $\psi_{(2)}$, and the superposition $\psi_{mix}$ of the second and the third eigenfunctions, individually. As expected, the behaviors of the bulk scalar field in the first two panels are qualitatively similar to that considered in the last subsection. The first panel shows that the scalar field initially evolves to the intermediate state with two crests, which indicates that the system initially evolves to the excited state with $n=1$. Then, the scalar field eventually reaches a state with only one crest, which implies that the system eventually evolves into the ground state with $n=0$. The second panel shows that the system initially evolves to the excited state with the overtone number $n=2$, and eventually reaches the ground state with the overtone number $n=0$. The third panel shows that the scalar field initially evolves to the intermediate state with three crests, and then evolves to the intermediate state with two crests, and eventually to the final stationary state with only one crest. This implies that the system evolves to the $n=2$ excited state, which then decays to the $n=1$ excited state, and finally to the $n=0$ ground state. By AdS/CFT, the evolution process of the system from the $n=2$ excited state to the $n=1$ excited state can be regarded as the non-equilibrium dynamical transition process between the excited states of the holographic superconductor.

\begin{figure}
\centering
  \includegraphics[width=8cm]{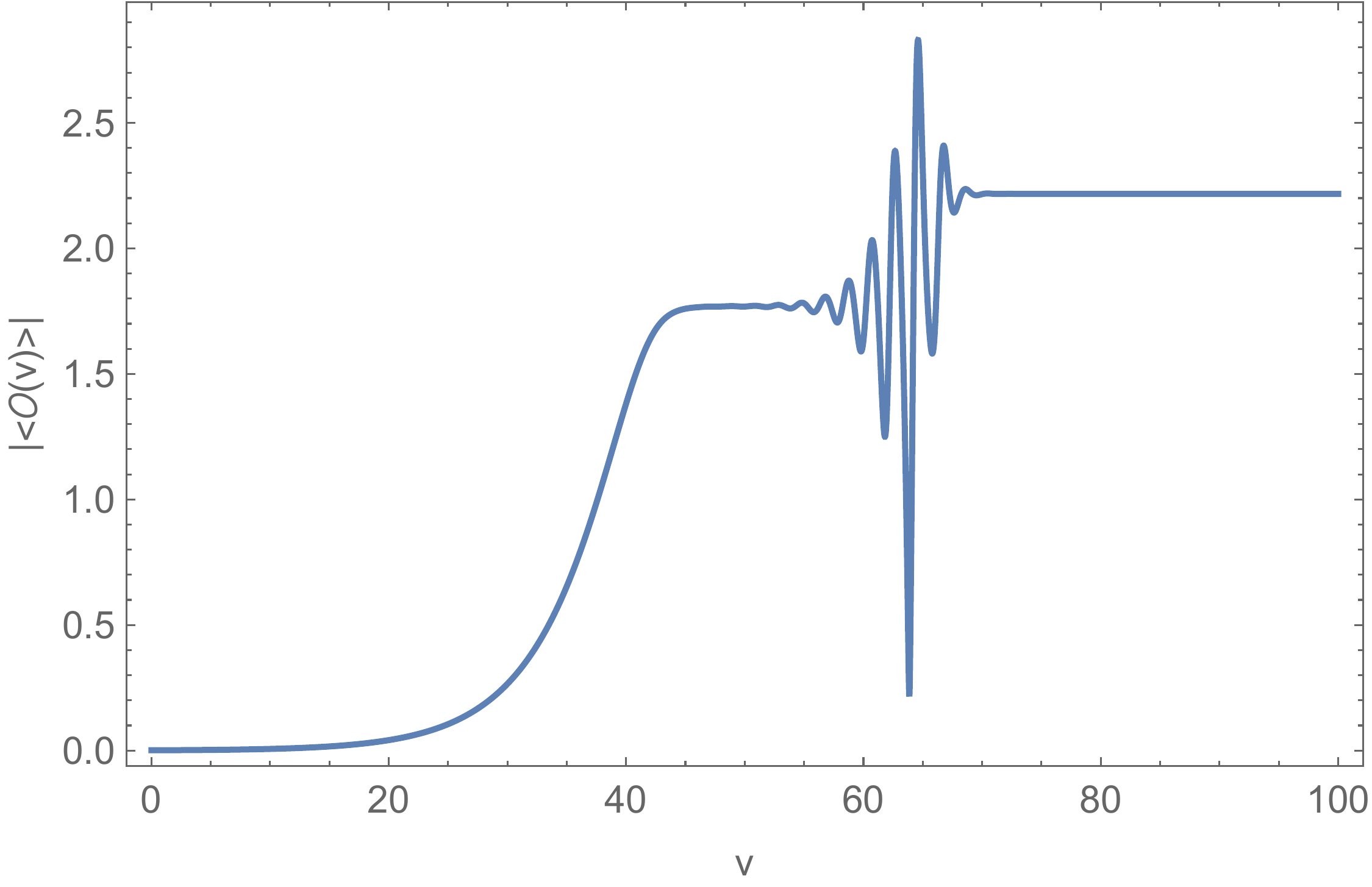}\\
  \includegraphics[width=8cm]{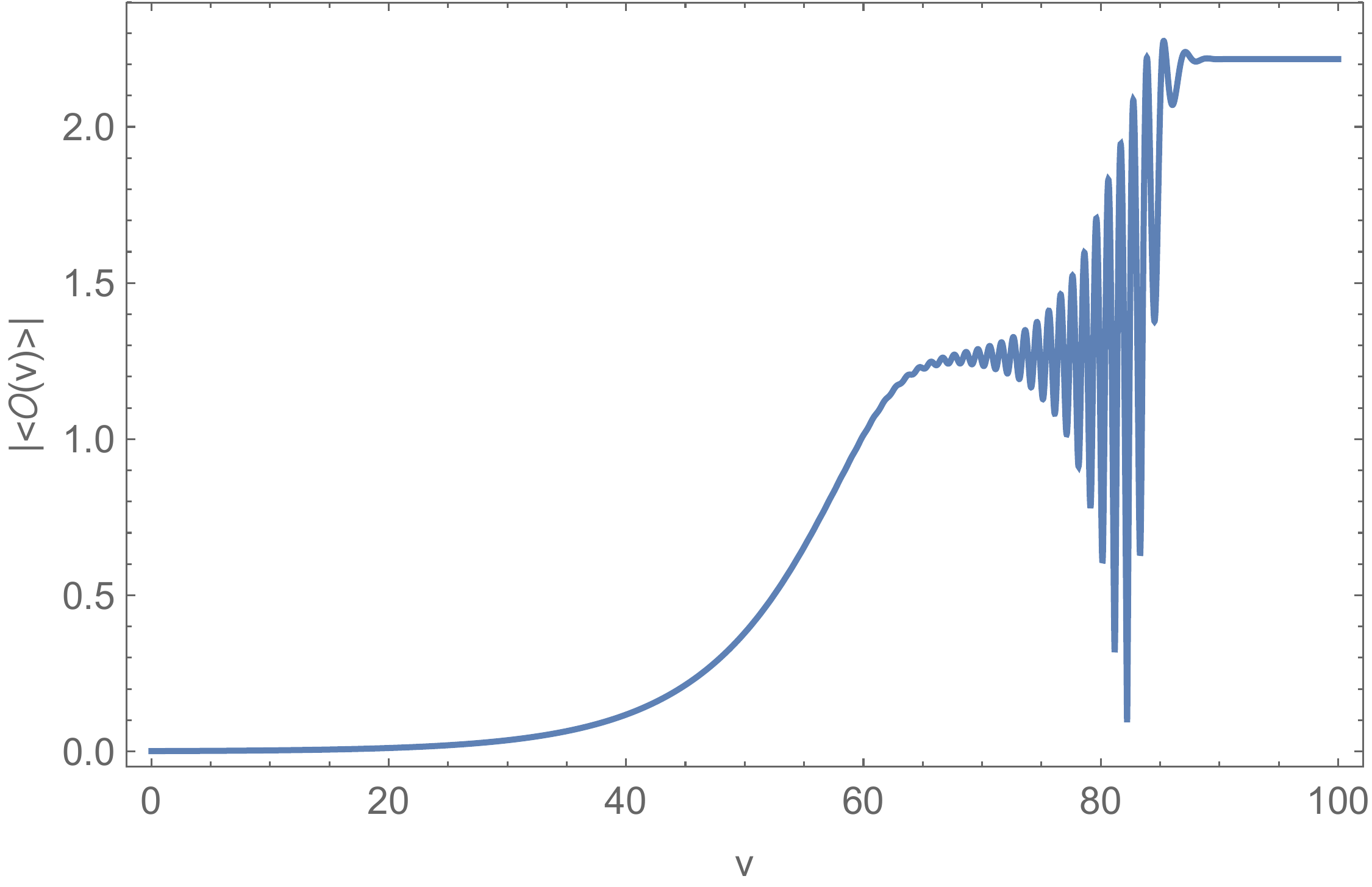}\\
  \includegraphics[width=8cm]{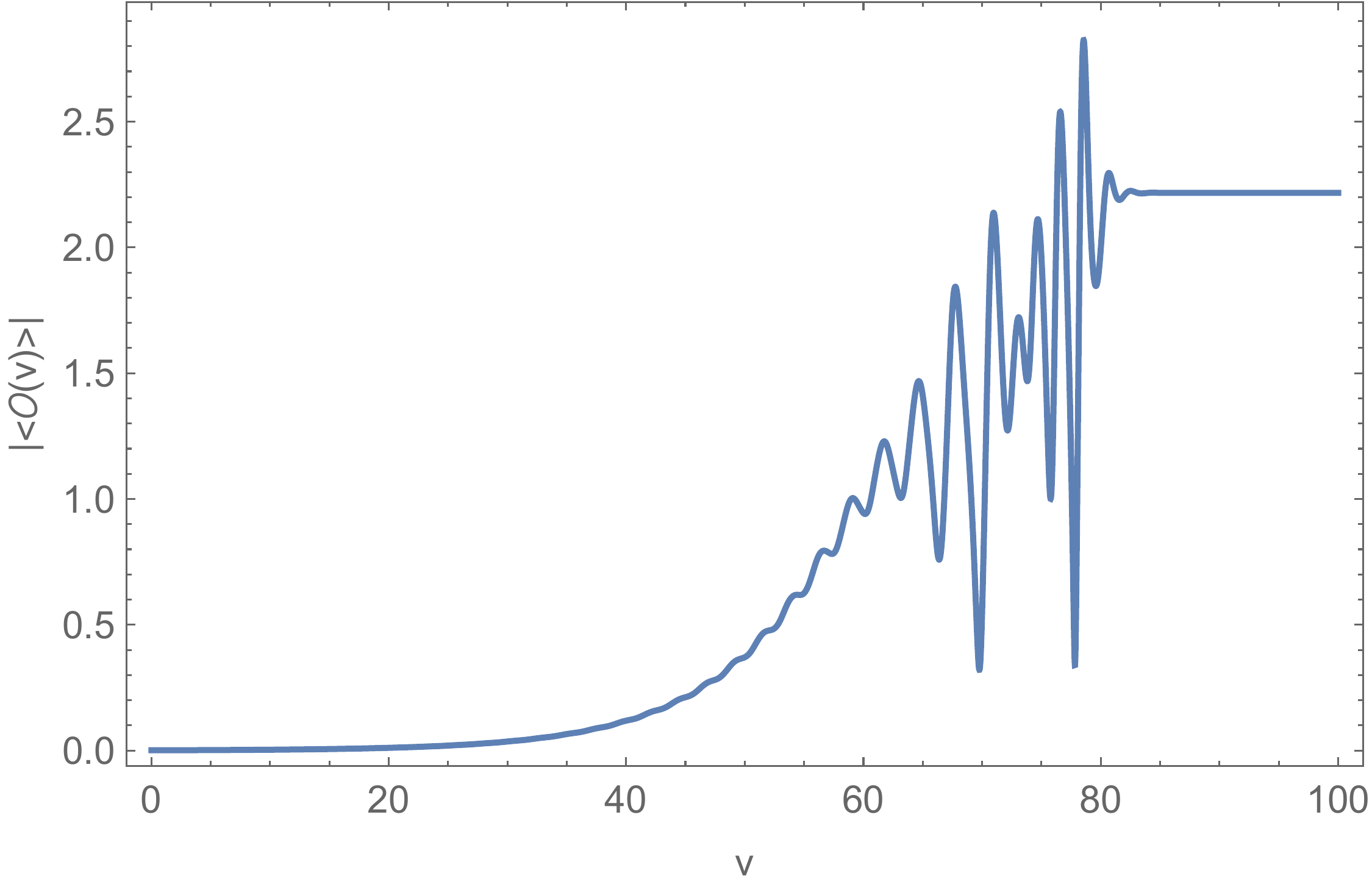}\\
  \caption{The time evolutions of superconducting order parameter $\langle \mathcal{O}\rangle$ for $Q=3.4$ and $q=6$ with different initial perturbations. In the three panels, the initial perturbation are taken as the second eigenfunction, the third eigenfunction, and the superposition of the second and the third eigenfunctions. }\label{OcaseII}
\end{figure}

The temporal profiles of the superconducting order parameters for the three cases are plotted in the three panels in Fig. \ref{OcaseII}, individually. As shown in the first panel, with the eigenfunction $\psi_{(1)}$ as the initial data in the evolution,  the time evolution of the order parameter is also qualitatively similar to that discussed in the last subsection. There is a plateau at the middle stage, which strongly indicates that the system reaches the intermediate state with $n=1$. The second panel is the time evolution of the order parameter with the eigenfunction $\psi_{(2)}$ as the initial data. At the first stage, the order parameter grows exponentially to reach the intermediate value, and then grows to the final value in an oscillating way. In this case, the system does not go through the $n=1$ excited state. As before, the oscillation behavior of the transition process from the $n=2$ excited state to the $n=0$ ground state is supposed to be induced by the interference between the $n=0$ and $n=2$ unstable modes. The corresponding oscillation frequency can be estimated as $|\textrm{Re}[\omega_0-\omega_2]|$, which is larger than the oscillation frequency $|\textrm{Re}[\omega_0-\omega_1]|$ in the first panel. This rough estimation is obviously consistent with the oscillating parts in the first two panels.

In the third panel of Fig. \ref{OcaseII}, the initial data is taken as the linear superposition of the eigenfunctions $\psi_{(1)}$ and $\psi_{(2)}$. The time evolution of the order parameter seems rather complicated. However, we can still observe that, at the first stage, the order parameter grows exponentially, which is determined by the single $n=2$ mode. Then, after oscillating twice, the order parameter reaches the constant value and does not change any more. This shows that the system initially evolves to the $n=2$ excited state, then decays to the $n=1$ excited state, and eventually reaches the $n=0$ ground state. The estimated oscillation frequency of the transition process from the $n=2$ excited state to the $n=1$ excited state, given by $|\textrm{Re}[\omega_1-\omega_2]|$, is smaller than the oscillation frequency $|\textrm{Re}[\omega_0-\omega_1]|$ from the $n=1$ state to the $n=0$ state, which is consistent with the whole behavior demonstrated in the third panel.

\begin{figure}
\centering
  \includegraphics[width=10cm]{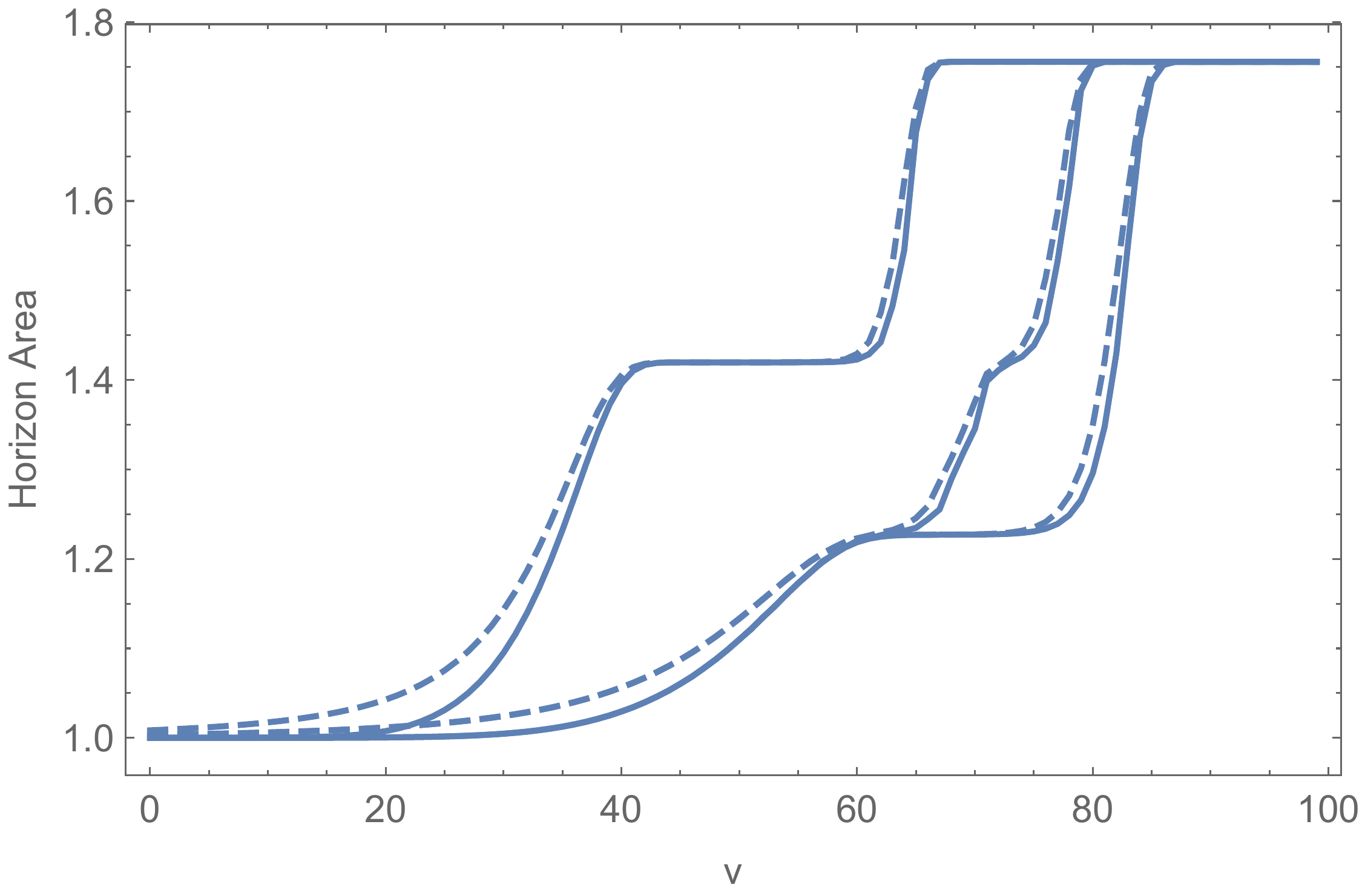}\\
  \caption{The time evolutions of the areas of horizons for $Q=3.4$ and $q=6$ with different initial perturbations. The dashed line represents the event horizon and the solid line represents the apparent horizon. The initial perturbations for the left two lines, the right two lines, and the middle two lines are the second eigenfunction, the third eigenfunction, and the superposition of the second and the third eigenfunctions, respectively. }\label{horizoncaseII}
\end{figure}

The areas of the event horizon and the apparent horizon are presented in Fig. \ref{horizoncaseII}.
from the middle two curves, we see a sharp signal that the system initially evolves to the $n=2$ excited hairy black hole, then decays to $n=1$ excited hairy black hole, and eventually reaches the $n=0$ ground state. At the early stage of the evolution, the middle two curves coincide with the right two curves. The reason is that there is a very small portion of the second eigenfunction $\psi_{(1)}$ in the initial data for the middles two curves. However, their behaviors are different after the system reaches the first $n=2$ intermediate state. The system with the single overtone mode $\psi_{(2)}$ as the initial data evolves to the final $n=0$ state directly, while the system with the mixed initial data $\psi_{mix}$ decays first to the $n=1$ excited state, and eventually decays to the final $n=0$ ground state. It can be observed that the time that the system stays at the intermediate state is rather short for the mixed initial data. This makes the dynamical transition process between the excited states very difficult to capture. Another observation is that, for the system with the fixed mass and charge, the horizon area of the excited hairy black hole with the larger overtone number is always smaller than the area of the hairy black hole with the smaller overtone number, where the bald RNAdS black hole can be regarded as the black hole with the infinite overtone number.

\begin{figure}
\centering
  \includegraphics[width=10cm]{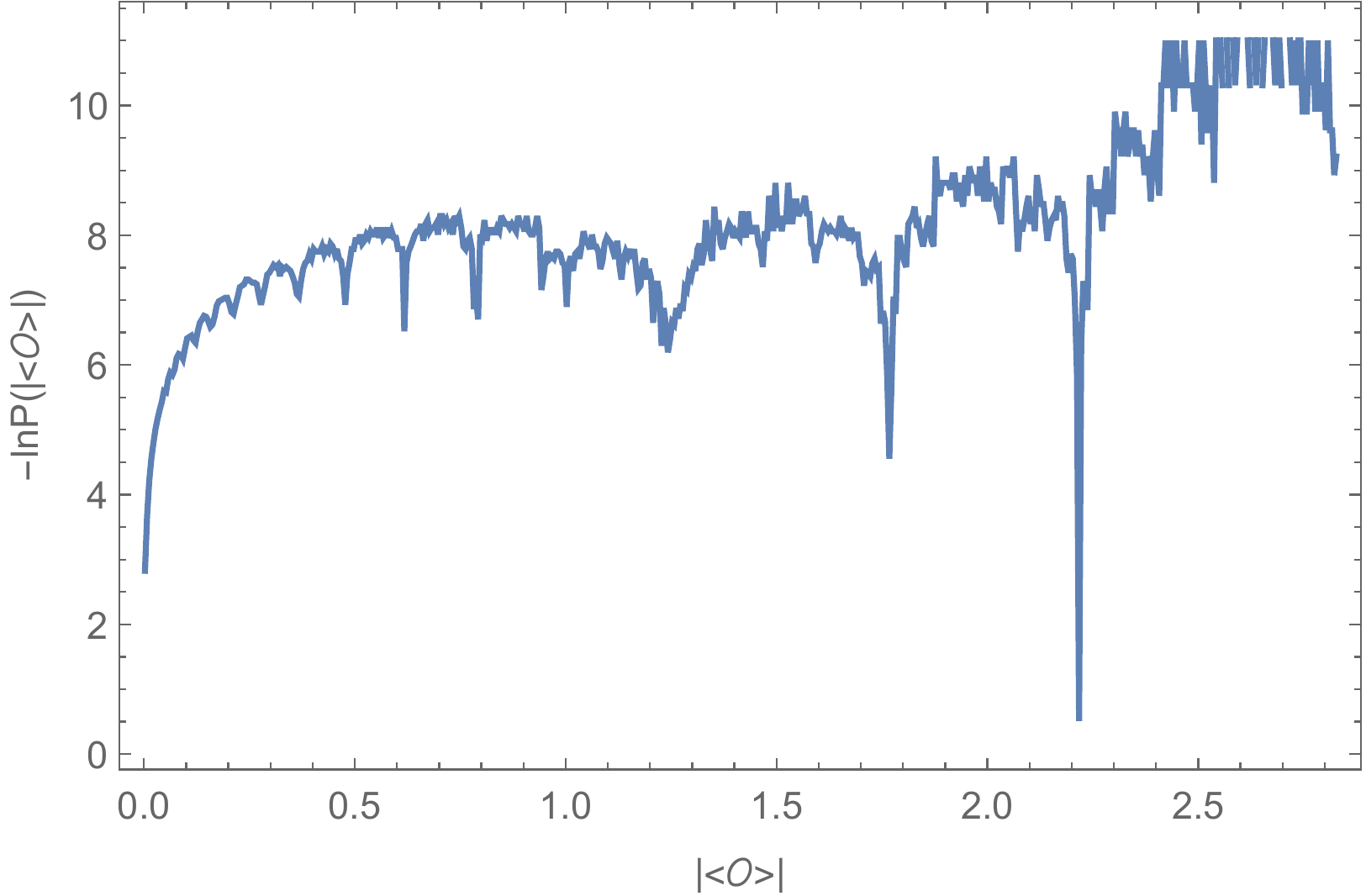}\\
  \caption{Probability distribution of the superconducting order parameter in the whole evolution process for $Q=3.4$ and $q=6$. Typically, the order parameter is equally discretized with the interval $\delta=0.005$.}\label{LnPcaseII}
\end{figure}

To analyze the effective landscape, the finite time trajectories with different initial conditions form the statistical ensemble and the corresponding data are gathered together. From the time traces of the order parameter $|\langle \mathcal{O}\rangle|$ under different initial data, we collect the statistics and obtain the the probability distribution $P(|\langle \mathcal{O}\rangle|)$ as a function of the superconducting order parameter $|\langle \mathcal{O}\rangle|$. We display the effective potential or free energy $F/kT\sim -\ln P$ in Fig. \ref{LnPcaseII}. It is clear that there are four wells at $|\langle \mathcal{O}\rangle|=0$, $1.24$, $1.76$, and $2.21$, individually, which implies that the system has higher chances of staying at the normal RNAdS black state, at the $n=2$, the $n=1$ excited state, and the $n=0$ ground state black holes. The depth of the well represents the time duration that the system stays at the different hairy black hole states in the bulk or the corresponding normal, ground, first, and second excited superconducting states on the boundary. We can observe that the weight or the duration that the system stays at the $n=2$ excited state is smaller or shorter than that at $n=1$ excited state. The effective landscape also implies that the ground state black hole is stable while the $n=1$ and the $n=2$ excited state black holes are less stable or metastable. If the evolution time approaches to infinity, the probability that the system stays at the ground state hairy black hole or the corresponding superconducting state approaches to one. The overall landscape is then funneled shaped towards the ground state black hole in the bulk or the superconducting ground state at the boundary with certain degrees of roughness, when the initial temperature is below the critical temperature $T_c$. The funneled landscape gives rise to a guiding force for the evolution from the bald RNAdS black hole through the 2nd and first excited states and finally to the ground state of the hairy black hole in the bulk or from the normal state through the second and first intermediate or excited states and finally to the superconducting ground state on the boundary. The landscape topography in terms of the barrier height among these wells can be used to quantify the likelihood and duration for the transitions among these states. The funneled landscape with roughness leads to a global physical picture not only for the quantification of the weight of each state, but also for characterizing the barrier crossing and cascade for the underlying dynamical evolution process towards the ground state.

\section{Conclusion and discussion}\label{end}

In summary, by numerically solving the bulk gravitational dynamics, we have studied the non-equilibrium condensation process of the holographic $s$-wave superconductor with the excited states as the intermediate states during the relaxation. When choosing the $n$th overtone eigenfunction of the linearized equation of motion of the scalar field as the initial perturbations in the nonlinear evolution, we have observed that the bulk system firstly evolves to metastable intermediate states that corresponds to the excited states of the holographic superconductors with the same overtone number on the AdS boundary. Generically, the system remain in such an excited state for an appreciable period. So one can harness this process to prepare the excited holographic $s$-wave superconductor. With the linear superposition of the eigenfunctions as the initial data, we have also observed the transition process between the excited states of the holographic superconductor. Because the intermediate state is metastable, the system eventually evolves to the final stationary state that corresponds to the ground state of the holographic superconductor. We also provide a global and physical picture of the evolution dynamics of the black hole and the corresponding superconducting phase transition from the funneled landscape view, quantifying the weights of the states and characterizing the transitions and cascades towards the ground state.

There are some related problems that should be investigated in the future. The first is how the non-equilibrium dynamics will be influenced by the anisotropy of the background black hole spacetime. The second question is to discuss the corresponding non-equilibrium dynamics of the holographic $p$-wave and $d$-wave superconductors at the excited states. In \cite{VP}, by employing the time-dependent Ginzburg-Landau equations, the transition processes between the metastable states of a mesoscopic superconducting ring are investigated in the presence of an external magnetic field. Therefore, another question is whether the system can be driven from the ground state to the excited state under some external source by using the holographic method. These questions deserve further investigations in the future.

\section*{Acknowledgement}

H.Z. is supported in part by NSFC with Grant NSFC 11675015, by FWOVlaanderen through the project G006918N, and by the Vrije Universiteit Brussel through
the Strategic Research Program ``High-Energy Physic''. He is also an individual FWO
fellow supported by 12G3518N.

\end{document}